\documentclass{emulateapj}

\newcommand\HII{H\,{\sc ii} }

\slugcomment{(May 26,2011; Accepted for publication in ApJL)}

\shorttitle{Continuum and Spectral Index imaging with the EVLA}
\shortauthors{Bhatnagar et al.}

\usepackage{natbib}
\begin{document}

\title{Extended Very Large Array Observations Of Galactic Supernova Remnants: Wide-Field Continuum
And Spectral-Index Imaging}

\author{S. Bhatnagar}
\affil{National Radio Astronomy Observatory, Socorro, NM - 87801, USA.}
\email{sbhatnag@nrao.edu}

\author{U. Rau}
\affil{National Radio Astronomy Observatory, Socorro, NM - 87801, USA.}
\email{rurvashi@nrao.edu}

\author{D. A. Green}
\affil{Astrophysics Group, Cavendish Laboratory, J J Thomson Ave., Cambridge CB3 0HE, U.K.}
\email{dag9@cam.ac.uk}

\and

\author{M. P. Rupen}
\affil{National Radio Astronomy Observatory, Socorro, NM - 87801, USA.}
\email{mrupen@nrao.edu}

\begin{abstract}
  The radio continuum emission from the Galaxy has a rich mix of
  thermal and non-thermal emission.  This very richness makes their
  interpretation challenging since the low radio opacity means that a
  radio image represents the sum of all emission regions along the
  line of sight.  These challenges make the existing narrow-band radio
  surveys of the Galactic plane difficult to interpret: e.g., a small
  region of emission might be a supernova remnant (SNR) or an \HII
  region, or a complex combination of both.  Instantaneous wide
  bandwidth radio observations in combination with the capability for
  high resolution spectral index mapping, can be directly used to
  disentangle these effects.  Here we demonstrate simultaneous
  continuum and spectral index imaging capability at the {\it full
    continuum sensitivity and resolution} using newly developed
  wide-band wide-field imaging algorithms.  Observations were done in
  the {\it L}- and {\it C}-Band with a total bandwidth of 1 and 2~GHz
  respectively.  We present preliminary results in the form of a
  full-field continuum image covering the wide-band sensitivity
  pattern of the EVLA centered on a large but poorly studied SNR
  (G$55.7+3.4$) and relatively narrower field continuum and spectral
  index maps of three fields containing SNR and diffused thermal
  emission.  We demonstrate that spatially resolved spectral index
  maps differentiate regions with emission of different physical
  origin (spectral index variation across composite SNRs and
  separation of thermal and non-thermal emission), superimposed along
  the line of sight.  The wide-field image centered on the SNR
  G$55.7+3.4$ also demonstrates the excellent wide-field wide-band
  imaging capability of the EVLA.
\end{abstract}

\section{Introduction}

In order to test the wide-band imaging capabilities of the
EVLA\footnote{The EVLA is operated by the National Radio Astronomy
  Observatory (NRAO) for Associated Universities Inc. under a license
  from the National Science Foundation of the USA.}, we have
observed several known and possible Galactic supernova remnants
(SNRs). There are 274 Galactic SNRs catalogued
\citep{SNRCatGreen2009}, the majority of which are classified as
`shell' remnants, showing limb-brightened synchrotron radio emission,
typically with $S \propto \nu^{\sim -0.5}$. Some SNRs are classed as
`filled-center' remnants (or `plerions'), as they show emission at
radio wavelengths that is brightest in the center, due to the power
input from a central pulsar. These filled-center remnants generally
have much flatter radio spectra at GHz-frequencies, with $S \propto
\nu^{\sim -0.0{-}0.3}$. Finally, there are some remnants that are
classed as composite remnants, as they show a mixture of
limb-brightened shell-like emission, with a filled-center-like core
(i.e.,\ a pulsar wind nebula) showing a much flatter spectrum. In
addition to the known remnants, there are many other possible and
probable Galactic SNRs that have been proposed, for which additional
observations are needed to confirm their nature. In many regions of
the Galactic plane, particularly near to $b=0^\circ$ and $l=0^\circ$,
radio emission from SNRs is easily confused with thermal radio
emission (typically with $S \propto \nu^{\sim -0.1}$) from {\sc H\,ii}
regions. The wide-band of the EVLA allows a variety of spectral
studies of SNRs to be made, in order to: (1) investigate spectral
variations across the face of shell remnants, in order to study the
differences in the shock acceleration mechanism at work in different
regions of the remnants; (2) study the flat spectrum cores of
composite remnants; and (3) clarify the nature of proposed possible
SNRs, by disentangling the non-thermal synchrotron emission from
confusing thermal emission.

We have observed two smaller known remnants G16.7$+$0.1 and
G21.5$-$0.9.  G16.7$+$0.1 is a `composite' remnant, $\sim 3\arcmin$
across (e.g.,\ \citealt{1989ApJ...341..151H}). G21.5$-$0.9 was classed
as a filled-center remnant, only $\sim 1.5\arcmin$ in extent, until a
faint X-ray shell $\sim 4\arcmin$ in diameter was identified by
\cite{2000ApJ...533L..29S}. Since then it has been classed as a
composite remnant, although a radio counterpart to the X-ray shell has
not been identified (see \citealt{2011MNRAS.412.1221B}).  G21.5$-$0.9
contains a pulsar, which has a high spin-down luminosity
(\citealt{2005CSci...89..853G}). We also observed a field centered at
$l=19.6^\circ$, $b=-0.2^\circ$, which contains `high-probability' SNR
candidate, G19.66$-$0.22, identified by \cite{HelfandMAGPIS}.
Finally, we observed one larger known but poorly studied shell
remnant, G55.7$+$3.4, which is $\sim 20 \times 25$~arcmin$^2$ in
extent (see \cite{Goss_G55.7+3.4}). An old, probably unrelated pulsar
lies within the northern edge of this remnant.

\section{Observations and Calibration}

The observations were made using the EVLA in D-array.  The {\it L}-Band
observations used the full 1~GHz bandwidth covering the frequency
range of $1-2$~GHz.  The {\it C}-Band observations used 2~GHz bandwidth in
two separate $\sim1$~GHz bands.  The observations parameters for all
the fields are listed in Table~\ref{Tab:ObsParams}. At the time of
observations, not all antennas in the array were outfitted with the
new {\it L}- and {\it C}-Band receivers.  Due to this and other failures during
the commissioning phase of the EVLA, typically 20 antennas were used
for these observations.  The {\it L}-Band data was recorded in 8 or 16
spectral windows (SPW) at a frequency resolution of 2 and 1~MHz
respectively with an integration time of 1~s.  The {\it C}-Band data
was recorded in 16 SPWs at a resolution of 2~MHz.  High time and
frequency resolution was required to allow effective removal of radio
frequency interference (RFI), accurate bandpass calibration and
corrections for wideband group delay present in the data from EVLA
WIDAR correlator.  To reduce the data volume, after initial data
flagging the data were averaged to 10~s resolution in time.  The
standard flux calibrators 3C286 and 3C147 were observed for flux
calibration and the compact sources J1822$-$0938 (for G16.7$+$0.1,
G19.6$-$0.2 and G21.5$-$0.9) and J1925$+$2106 (for G55.7$+$3.4) were
observed at intervals of 30~min for phase and bandpass calibration.
Using known frequency dependent model for the flux calibrators
(including spectral index) antenna gains for each frequency channel
across the observed bandwidth were determined for flux and group delay
calibration.  The antenna gains for each SPW were also determined
using the phase calibraters.  In order to account for the spectral
index of the calibraters, a model including the frequency dependent
flux was determined using the multi-scale multi-frequency synthesis
(MS-MFS) imaging algorithm which simultaneously makes Stokes-{\it I} and
spectral index maps.  The final complex bandpass shape was determined
using this model and applied to the data for the target fields.  This
calibration procedure allowed calibration of time-varying direction
independent gains without transferring the effects of the frequency
dependent flux of the calibraters to the target fields.

\section{Wide-band Wide-field Imaging}
RFI affected frequency channels were flagged throughout the observing
band.  Full SPWs covering the frequency range $1.516 \sim
1.643$~GHz had to be dropped due to the presence of strong RFI at
{\it L}-Band.  Ten channels at the edges of the SPWs were also flagged due
to the roll-off of the digital filters in the signal chain. The rest
of the frequency channels were used for wide-band imaging.

Conventional image reconstruction techniques for interferometric
imaging fundamentally models the brightness distribution as a
collection of scaleless components (amplitude per pixel) leaving
errors which limit the imaging performance.  Scale-sensitive methods
\citep{Asp_Clean,MSCLEAN} significantly reduce the magnitude of such
errors by explicitly solving for the scale size of emission across the
field of view.  Imaging performance using wide-band interferometric
observations of fields with extended emission is additionally limited
by the fact that not only does the scale of emission vary across the
field of view, the spectral properties also vary across the field.
The dominant source of these spectral variations is the change of
antenna primary beam with frequency and spatial dependence of the
spectral index of the radio emission.  Both effects result in
variation of the strength of emission as a function of frequency in a
direction dependent manner.  Furthermore, antenna primary beams are
typically rotationally asymmetric which result in time-varying gains
due to the rotation of the primary beams with parallactic angle for
El-Az mount antennas.  Scale-sensitive imaging reconstruction
algorithms that can also {\it simultaneously} account for time and
frequency-dependent effects are therefore required for wide-band
wide-field imaging with the EVLA, particularly in the Galactic plane.
For this, two new algorithms, namely the MS-MFS
\citep{RAU_THESIS,MSMFCLEAN} and the A-Projection \citep{AWProjection}
algorithm, have been developed.  The MS-MFS algorithm combines the
scale-sensitive deconvolution with explicit frequency dependent
modeling of the emission to map the spatial distribution of the
spectral index in a scale-sensitive manner.  The A-Projection
algorithm on the other hand corrects for the time and frequency
dependence of the antenna primary beams as part of the iterative image
reconstruction.

We used the MS-MFS \citep{RAU_THESIS,MSMFCLEAN} algorithm implemented
in the $\mathcal{CASA}$\footnote{{\tt http://casa.nrao.edu}}
post-processing package, for wide-band imaging.  The spectral index
information derived using the traditional method of using images made
at different frequencies make spectral index images at the {\it
  lowest} resolution and at the sensitivity of {\it single} channel
bandwidth. Spectral index images made using the MS-MFS algorithm on
the other hand does not require smoothing the images to the lowest
resolution and produces the spectral index images at the full
sensitivity and resolution offered by the data.  This must be combined
with techniques to correct for various time variable PB-effects during
image deconvolution to achieve noise limited imaging performance.
While the A-Projection algorithm can be used to correct for these
PB-effects during image deconvolution, for the images presented here
we used only the time-averaged PBs ({see Figure~\ref{Fig:AvgPB}}) to
apply correction post deconvolution.  Work towards a fully integrated
wide-field wide-band imaging algorithms is in progress.

\begin{table*}
\begin{center}
\caption{List of Observations Parameters}
\begin{tabular}{clccl}
%\tablewidth{0pt}
\tableline\tableline
Field & Pointing Center & Date & Frequency Range & Calibrators \\
      & (J2000)         &      &   (GHz)         & Flux, Phase \\
\tableline
G16.7$+$0.1 & RA=18:20:57  & 2010 Aug 12  & 4.49--5.45 & J1331$+$3030 and J0137$+$331,\\
\nodata     & Dec=-14:19:30 & \nodata    & 6.89--7.85\tablenotemark{a} & J1822$-$0938  \\
G19.6$-$0.2 & RA=18:27:38  & 2010 Jul 19 & 1.00--2.03   & J1331$+$3030,  J1822$-$0938 \\
\nodata     & Dec=-11:56:32 & \nodata    & \nodata    & \\
G21.5$-$0.9 & RA=18:33:32  & 2010 Aug 12  & 4.49--5.45 & J1331$+$3030 and J0137$+$331, \\
\nodata     & Dec=-10:34:10 & \nodata    & 6.89--7.85\tablenotemark{a} & J1822$-$0938  \\
G55.7$+$3.4 & RA=19:21:40  & 201 Aug 23  & 1.00--2.03   & J1331$+$3030 and J0542$+$498,\\
\nodata     & Dec=+21:45:00 & \nodata    & \nodata    &  J1925$+$2106\\
\tableline
\end{tabular}
\tablenotetext{a}{The two ranges correspond to the two 1-GHz wide
  frequency bands covering total of 2~GHz bandwidth at {\it C}-Band.}
\label{Tab:ObsParams}
\end{center}
\end{table*}

\section{Results}
\begin{figure*}
\centering
\vbox{
  \hbox{
    \includegraphics[width=6.2cm]{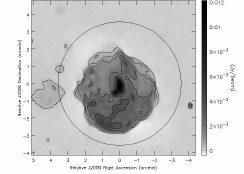}
    \includegraphics[width=6.0cm]{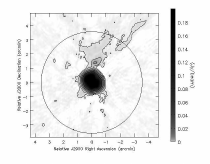}
    \includegraphics[width=6.0cm]{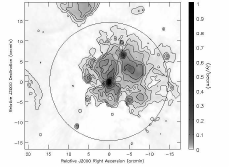}
  }
  \hbox{
    \includegraphics[width=6.2cm]{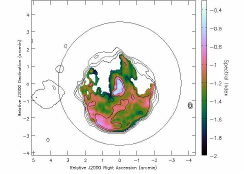}
    \includegraphics[width=6.2cm]{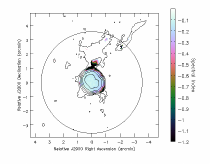}
    \includegraphics[width=6.0cm]{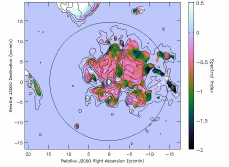}
  }
}
\caption{Stokes-{\it I} (top row) and Spectral Index images
  (bottom row).  The first two columns show {\it C}-Band images of the SNRs
  G16.7$+$0.1 and G21.5$-$0.9 respectively.  The resolution in these
  images is 10\arcsec\ and RMS noise is $11\mu Jy~beam^{-1}$ and
  $30\mu~Jy~beam^{-1}$ respectively. The last column shows the {\it L}-Band image
  of G19.6$-$0.2 with a compact source of thermal emission of positive
  spectral index surrounded by non-thermal emission with a negative
  spectral index.  The resolution in this image is 40\arcsec\ and RMS
  noise of $0.2m~Jy~beam^{-1}$.  The circle indicates the half-power point of
  the antenna primary beam at the reference frequencies of 6.0~GHz for
  {\it C}-Band images and 1.5~GHz for the {\it L}-Band image.}
\label{Fig:SNRs_NarrowField}
\end{figure*}

\subsection{G16.7$+$0.1, G21.5$-$0.9, G19.6$-$0.2}
Figure~\ref{Fig:SNRs_NarrowField} shows the Stokes-{\it I} and Spectral
Index images of two catalogued SNRs (G16.7$+$0.1 and G21.5$-$0.9)
\citep{SNRCatGreen2009} and one candidate SNR (G19.6$-$0.2).  The top
row shows the Stokes-{\it I} images and the bottom row shows the spectral
index images, with the Stokes-{\it I} contours overlaid on the spectral
images.  Correction for the primary beam response raises the noise
away from the center.  In order to show the detailed brightness
distribution, we chose to show the Stokes-{\it I} images without correcting
for the primary beam response.  The integrated flux reported below was
determined from PB-corrected images.  The spectral index images have
been corrected for the average primary beam effects, since the effect
of primary beam response in the spectral index images is stronger.
Image of all the fields covering the full wide-band sensitivity show
many more compact and diffused sources.  Due to space limitation, here
we show central portions of these images.

The first column has the images at a reference frequency of 6~GHz, of
the composite SNR G16.7$+$0.1 at a resolution of 10\arcsec.  The SNR
has a flatter spectrum core with an average spectral index of
$-0.54\pm0.02$ ($S \propto \nu^\alpha$) surrounded by a relatively
steeper spectrum nebula with the spectral index ranging from $-0.8$ to
$1.1$.  The total continuum integrated flux is 1.2~Jy with an RMS
noise of 11~$\mu~Jy~beam^{-1}$.  The diffused sources on east of SNR are
unrelated but real sources of emission.  The images in the second
column are of the filled-centered SNR G21.5$-$0.9, also at the
reference frequency of 6~GHz.  The RMS noise in this image is
$\sim30\mu~Jy~beam^{-1}$ and the resolution is 10\arcsec.  The spectral
index across this SNR is uniform across with a value of $+0.12\pm
0.03$ and integrated flux density of 6~Jy.  Also visible in the image
is the compact source called the ``northern knot'' about 2\arcmin\ to
the north.  The long extended feature extending towards the north-west
direction is part of the separate catalogued SNR G21.6$+$0.84.  The
third column contains the {\it L}-Band images at a reference frequency of
1.5~GHz, of the region centered on the Galactic co-ordinates
$l=19.6^\circ, b=-0.2^\circ$.  This field contains a candidate SNR
G19.66$-$0.22 \citep{HelfandMAGPIS}.  Stokes-{\it I} image shown in the top
row shows weak diffused emission surrounding two compact strong
sources, making it difficult to determine the nature of the objects
based on Stokes-{\it I} morphology alone.  The spectral index image however
clearly separates sources of thermal and non-thermal emission.  The
spectral index of the diffused emission range from $-0.4$ to $-0.5$.
The spectral index of the compact source (G19.67$-$0.15) towards the
norther edge of the diffused nebula has a spectral index of $-0.3$.
The superimposed compact thermal source near the center is the known
ultra-compact \HII region G19.61$-$0.23
(e.g., \cite{UCHII_MORPHOLOGIES,
  Hofner_Churchwell_1996A&AS..120..283H}) also with a spectral index
of $+0.3$.  Based on morphology of the spectral distribution we
conclude that this field contains an SNR, possibly of filled-centered
type.

\subsection{G55.7$+$3.4}

\begin{figure*}
%\hbox{
%  \includegraphics[width=17cm]{figure_G55p7_bw_big.png}
%  \includegraphics[width=17cm]{f2.eps}
  \includegraphics[width=17cm]{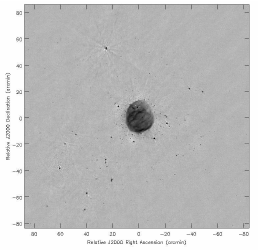}
%}
  \caption{Stokes-{\it I} image at a reference frequency of
    1.5~GHz.  Data in the frequency range 1.256--1.905~GHz were
    included for this image.  For this frequency range, the HPBW of
    the primary beam changes from 32\arcmin\ to 21\arcmin. This
    $2^\circ \times 2^\circ$ image covers the full-field corresponding
    to the wide-band sensitivity pattern of the EVLA ({see
    Figure~\ref{Fig:AvgPB}}) of the field containing the SNR G55.7+3.4.}
\label{Fig:G55.7+3.4}
\end{figure*}

\begin{figure*}
\hbox{
  \includegraphics[width=9cm]{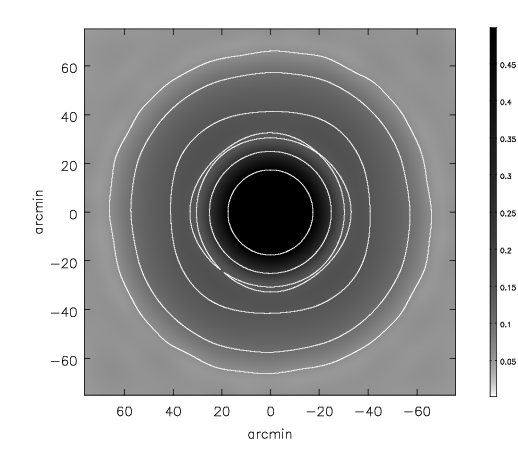}
  \includegraphics[width=9cm,height=8.2cm]{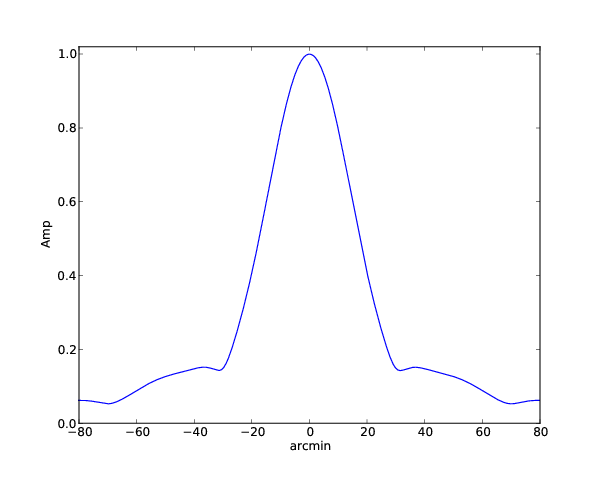}
}
\caption{The EVLA time-averaged wide-band sensitivity
  pattern covering the same frequency range and the area shown in
  Figure~\ref{Fig:G55.7+3.4}.  The left panel shows the pattern in gray
  scale with overlaid contours at 50\%, 25\%, 15\%, 10\% and 6\% of
  the peak value.  The right panel shows a one-dimentional cut through
  the pattern shown on the left.}
\label{Fig:AvgPB}
\end{figure*}

The SNR G55.7+3.4 is classified as an ``incomplete-shell SNR'' with a
pulsar within the boundary of the remnant based on the only existing
radio image at 610 and 1.415~GHz by \cite{Goss_G55.7+3.4}.  These are
the only image published in the literature of this SNR made over three
decades ago using the Westerbork Synthesis Radio Telescope.  The full
$2^\circ \times 2^\circ$ Stokes-{\it I} {\it L}-Band image centered on
this SNR at the reference frequency of 1.5~GHz covering the full
wide-band sensitivity pattern of the EVLA antenna is shown in
Figure~\ref{Fig:G55.7+3.4}.  This is the highest resolution and
sensitivity image of this region to date.  The wide-band sensitivity
pattern of the EVLA extended significantly beyond even the first
sidelobe of the PB at the lowest frequency in the band.  {The
  time-averaged wide-band sensitivity pattern of the EVLA for the
  equivalent frequency range and field-of-view is shown in the left
  panel of Figure~\ref{Fig:AvgPB}.  The one-dimensional slice through
  this pattern in the right panel, shows the sensitivity pattern
  extending significantly beyond the ``main-lobe'' at the level of
  $5-15\%$ of the peak value.}  With sources, some strong, detected up
to $1^\circ$ away from the center of the image, imaging the full field
was required to reach noise-limited imaging at the center of the
field.  Due to this wide field of view, even for D-array observations,
correction for the effects of non-coplanar baselines had to be
applied.  This was done using the W-Projection algorithm
\citep{WProjection}.  The Stokes-{\it I} image of the SNR shows a
nearly complete shell-type SNR.  At a resolution of 30\arcsec, this
image for the first time, shows a network of filamentary structures
along with smoother diffused emission filling the boundary of the
shell.  The spectral index image of this SNR however was not reliable
at all scales since the large scale emission is not well constrained
by the shortest measured baselines at all frequencies.  The filaments
however show up as regions of steeper spectral index compared to an
average spectral index of $-0.6$ of the surrounding regions.  The
strong compact source (G55.78$+$3.5) on the northern edge of the shell
in the Stokes-{\it I} image is an older unrelated pulsar (PSR
J1921$+$2153).  In the spectral index map (not shown here), this
pulsar shows up as a steep spectrum compact source with a spectral
index of $-2.2$.  The integrated flux of the shell, including the
pulsar and other foreground or background sources superimposed across
the shell is 1.0~Jy. The rms noise in the image is
10~$\mu Jy~beam^{-1}$.

\section{Discussion}

The technical goals of this pilot project were to assess our
low-frequency wide-band imaging capabilities for targets in the
Galactic plane. At {\it L}-Band and {\it C}-band, the Galaxy is replete with
compact and diffuse emission of both thermal and non-thermal
origin. Sources often have a mixture of thermal and non-thermal
emission, resulting in spectral indices that vary with position across
a given source and with spatial scale.  Conventional imaging
algorithms that rely on multiple narrow-band images to derive
spectral-index information are limited to the angular resolution
allowed by the lowest measured frequency, and the single-channel
sensitivity. The MS-MFS algorithm was designed to optimally use the
multi-frequency measurements to do spectral-index mapping across
diffused sources {\it at the highest sensitivity and angular
  resolution} offered by the data.

The following is a summary of our current status, based on the imaging
results shown in the previous section.  Our targets were chosen based
on existing information about their spatial and spectral structure, in
order to test the analysis methods in preparation for a Galactic plane
survey.

\begin{enumerate}

\item Separation of thermal versus non-thermal; steep versus steeper
  spectra : we see clear morphological separations between regions
  expected to have different spectral structure from the
  surroundings. These differences range from
  $\triangle\alpha\approx0.6$ between thermal and non-thermal regions,
  down to $\triangle\alpha\approx0.2$ between structures thought to be
  cores and shells. In regions with high S/N ratios ($>100$), the
  estimated uncertainty in the spectral index is $\pm0.03$.  In
  regions with low signal-to-noise ($<100$), the uncertainty in the
  spectral index rises rapidly, and all the images presented here used
  a Stokes-{\it I} S/N threshold of 100.

\item Image-reconstruction errors : errors in the spectral
  reconstructions are dominated by multi-scale deconvolution artifacts
  that decrease with appropriate choices of multi-scale imaging
  parameters (a set of scale-sizes with which to model the spatial
  structure). {Error estimates from Monte-Carlo simulations (by
    varying the choice of spatial scales) range from $\pm0.02$ when
    scales are chosen appropriately, up to 0.2 in the extreme case
    where diffuse emission is modeled by a series of
    $\delta$-functions.  These simulations were done for the EVLA
    D-configuration and the errors depend on the precise uv-coverage
    just as they depend for the standard imaging algorithms.  The
    MS-MFS algorithm reconstructs the frequency dependence of the flux
    by a polynomial fit across the frequency range on a per component
    basis.  The errors on the estimated spectral index therefore also
    depend on the errors of the fitted coeffcients.  These errors
    depend on variations in S/N ratio (across the
    frequency range or due to the scale of the emission) in the same
    way as the errors on the coefficients of any least-squares
    polynomial fitting algorithm (see
    ~\cite{Numerical_Recipes,MSMFCLEAN} for further details)}.
  Further, at the largest spatial scales sampled by the interferometer
  {(which, of course, is telescope dependent)}, the spectrum is
  unconstrained by the data and the MS-MFS model does not provide
  adequate additional constraints. Out of our examples, G55.7$+$3.4
  and G16.7$+$0.1 contain structure at spatial scales that fall within
  the unsampled central region of the interferometer sampling
  function, aleading to overall systematic errors in the spectral
  index across the source. These errors are limited to the largest
  scales, but in the current implementation of the algorithm, it is
  not possible to analyse the scales separately. In our examples,
  G55.7$+$3.4 is affected significantly by this error and spectra of
  only the most compact emission can be trusted.  Finally, in addition
  to errors in the spectral index, off-source artifacts are present in
  the Stokes-{\it I} images beyond dynamic-ranges of a few thousand
  (measured as the ratio of the peak brightness to the peak residual
  near the source). Dynamic-ranges for the images presented here range
  from 2000 for G16.7$+$0.1 to 10000 for G19.6$-$0.2. In most cases,
  wide-band self-calibration was required to achieve these limits.

\item Wide-field wide-band sensitivity of the instrument, and related
  errors : across a 2:1 bandwidth, the size of the antenna
  power-pattern changes enough to produce an average beam that is
  significantly different from the beams at individual
  frequencies. The most-prominent difference is the near lack of a
  first null, with sensitivity at the few percent level out to almost
  three times the HPBW. The G55.7$+$3.4 field is an example of this
  wide-field sensitivity pattern (the reference-frequency primary beam
  has a HPBW of 30\arcmin).  For sources far from the phase-center,
  the $w$-term causes the dominant wide-field error, and these images
  (in particular G55.7+3.4) are a result of W-Projection combined with
  MS-MFS.  The next dominant error (in Stokes-{\it I} and spectral index) is
  due to the frequency dependence of the primary beam. All the
  spectral-index maps shown here have been corrected for this effect
  using the time-averaged PB-model, and can be trusted out to the 25\%
  point of the reference-frequency primary-beam.  The time-variability
  of the primary-beams has not been accounted for, and is thought to
  be the cause of the next dominant error (visible as artifacts around
  bright sources about $1^\circ$ from the phase center in
  G55.7$+$3.4).

\end{enumerate}

The next steps in our analysis include combining multi-configuration
and multi-band data for the G16.7$+$0.1 and G19.6$-$0.2 fields to
introduce more constraints during image-reconstruction, to complete
the software integration required to combine MS-MFS and W-Projection
with the A-Projection algorithm to correct for time-varying
primary-beams, and further tests with mosaicking observations and how
the additional information aids the image reconstruction. These steps
are in accordance with our long-term goal of doing a wide-band
mosaiced survey of the Galactic plane at low frequencies.

\acknowledgments

We used the ATNF Pulsar Catalogue \citep{PSRCat2005}, accessible via
the URL {\tt http://www.atnf.csiro.au/people/pulsar/psrcat/} for the
co-ordinates of the pulsar in the G55.7$+$3.4 field.

% \bibliographystyle{apj}
% \bibliography{bhat0415}

\begin{thebibliography}{0}
\expandafter\ifx\csname natexlab\endcsname\relax\def\natexlab#1{#1}\fi

\end{thebibliography}


\begin{thebibliography}{17}
\expandafter\ifx\csname natexlab\endcsname\relax\def\natexlab#1{#1}\fi

\bibitem[{{Bhatnagar} \& {Cornwell}(2004)}]{Asp_Clean}
{Bhatnagar}, S., \& {Cornwell}, T.~J. 2004, Astron. \& Astrophys., 426, 747

\bibitem[{{Bhatnagar} {et~al.}(2008){Bhatnagar}, {Cornwell}, {Golap}, \&
  {Uson}}]{AWProjection}
{Bhatnagar}, S., {Cornwell}, T.~J., {Golap}, K., \& {Uson}, J.~M. 2008, Astron.
  \& Astrophys., 487, 419

\bibitem[{{Bietenholz} {et~al.}(2011){Bietenholz}, {Matheson}, {Safi-Harb},
  {Brogan}, \& {Bartel}}]{2011MNRAS.412.1221B}
{Bietenholz}, M.~F., {Matheson}, H., {Safi-Harb}, S., {Brogan}, C., \&
  {Bartel}, N. 2011, \mnras, 412, 1221

\bibitem[{{Cornwell}(2008)}]{MSCLEAN}
{Cornwell}, T.~J. 2008, IEEE Journal of Selected Topics in Signal Processing,
  issue 5,, 2, 793

\bibitem[{{Cornwell} {et~al.}(2008){Cornwell}, {Golap}, \&
  {Bhatnagar}}]{WProjection}
{Cornwell}, T.~J., {Golap}, K., \& {Bhatnagar}, S. 2008, IEEE JSTSP, 2, 647

\bibitem[{{Goss} {et~al.}(1977){Goss}, {Schwartz}, {Siddesh}, \&
  {Weiler}}]{Goss_G55.7+3.4}
{Goss}, W.~M., {Schwartz}, U.~J., {Siddesh}, S.~G., \& {Weiler}, K.~W. 1977,
  Astron. \& Astrophys., 61, 93

\bibitem[{{Green}(2009)}]{SNRCatGreen2009}
{Green}, D.~A. 2009, Bulletin of the Astronomical Society of India, 37, 45

\bibitem[{{Gupta} {et~al.}(2005){Gupta}, {Mitra}, {Green}, \&
  {Acharyya}}]{2005CSci...89..853G}
{Gupta}, Y., {Mitra}, D., {Green}, D.~A., \& {Acharyya}, A. 2005, Current
  Science, 89, 853

\bibitem[{{Helfand} {et~al.}(2006){Helfand}, {Becker}, {White}, {Fallon}, \&
  {Tuttle}}]{HelfandMAGPIS}
{Helfand}, D.~J., {Becker}, R.~H., {White}, R.~L., {Fallon}, A., \& {Tuttle},
  S. 2006, Astron. J., 131, 2525

\bibitem[{{Helfand} {et~al.}(1989){Helfand}, {Velusamy}, {Becker}, \&
  {Lockman}}]{1989ApJ...341..151H}
{Helfand}, D.~J., {Velusamy}, T., {Becker}, R.~H., \& {Lockman}, F.~J. 1989,
  \apj, 341, 151

\bibitem[{{Hofner} \&
  {Churchwell}(1996)}]{Hofner_Churchwell_1996A&AS..120..283H}
{Hofner}, P., \& {Churchwell}, E. 1996, Astron. \& Astrophys. Suppl. Ser., 120,
  283

\bibitem[{{Manchester} {et~al.}(2005){Manchester}, {Hobbs}, {Teoh}, \&
  {Hobbs}}]{PSRCat2005}
{Manchester}, R.~N., {Hobbs}, G.~B., {Teoh}, A., \& {Hobbs}, M. 2005, \aj, 129,
  1993

\bibitem[{{Press} {et~al.}(2007){Press}, {Teukolsky}, {Vetterling}, \&
  {Flannery}}]{Numerical_Recipes}
{Press}, W.~H., {Teukolsky}, S.~A., {Vetterling}, W.~T., \& {Flannery}, B.~P.
  2007, Numerical Recipes (3rd Edition), The Art of Scientific Computing
  (Cambridge University Press)

\bibitem[{{Rau}(2010)}]{RAU_THESIS}
{Rau}, U. 2010, PhD thesis, The New Mexico Institute of Mining and Technology,
  Socorro, New Mexico, USA

\bibitem[{{Rau} \& {Cornwell}(submitted,2011)}]{MSMFCLEAN}
{Rau}, U., \& {Cornwell}, T.~J. 2011,Astron. \& Astrophys., (accepted)

\bibitem[{{Slane} {et~al.}(2000){Slane}, {Chen}, {Schulz}, {Seward}, {Hughes},
  \& {Gaensler}}]{2000ApJ...533L..29S}
{Slane}, P., {Chen}, Y., {Schulz}, N.~S., {Seward}, F.~D., {Hughes}, J.~P., \&
  {Gaensler}, B.~M. 2000, \apjl, 533, L29

\bibitem[{{Wood} \& {Churchwell}(1989)}]{UCHII_MORPHOLOGIES}
{Wood}, D. O.~S., \& {Churchwell}, E. 1989, AJ Supp., 69, 831

\end{thebibliography}

% \begin{thebibliography}{}
% \end{thebibliography}

\end{document}